\documentstyle[preprint,aps,epsf]{revtex}
\newcommand{\beq}{\begin{equation}}
\newcommand{\eeq}{\end{equation}}
\newcommand{\ds}{\displaystyle}
\begin{document}
\draft
\tightenlines

\title  {
 Microscopic study of
 energy and centrality dependence of transverse collective flow in
 heavy-ion collisions
         }
\author {
L.V.~Bravina,$^{1,2}$ Amand Faessler,$^{1}$ C.~Fuchs,$^{1}$ 
and E.E.~Zabrodin$^{1,2}$ \\ 
  }
\address{$^1$ 
 Institute for Theoretical Physics, University of T\"ubingen,\\
 Auf der Morgenstelle 14, D-72076 T\"ubingen, Germany
         }
\address{$^2$ 
 Institute for Nuclear Physics, Moscow State University,
 119899 Moscow, Russia
         }

\maketitle

\begin{abstract}
The centrality dependence of directed and elliptic flow in light 
and heavy systems of colliding nuclei is studied within two 
microscopic transport models at energies from 1{\it A} GeV to 
160{\it A} GeV. The pion directed flow has negative slope in the 
midrapidity range irrespective of bombarding energy and mass number 
of the colliding ions. In contrast, the directed flow of nucleons 
vanishes and even develops antiflow in the midrapidity range in 
(semi)peripheral collisions at energies around 11.6{\it A} GeV and 
higher. The origin of the disappearance of flow is linked to nuclear 
shadowing. Since the effect is stronger for a light system, it can be 
distinguished from the similar phenomenon caused by the quark-gluon 
plasma formation. In the latter case the disappearance of the flow due 
to the softening of the equation of state should be most pronounced in 
collisions of heavy ions. The centrality dependence of the elliptic 
flow shows that the maximum in the $\langle v_2(b) \rangle $ 
distribution is shifted to very peripheral events with rising incident 
energy, in accord with experimental data. 
This is an indication of the transition from baryonic to mesonic 
degrees of freedom in hot hadronic matter.
\end{abstract}


\widetext

\section{Introduction}
\label{sec1}

Collective effects, such as the expansion of highly compressed 
nuclear matter in the direction perpendicular to the beam axis of 
colliding heavy ions at relativistic energies, are very important 
for the study of the nuclear equation of state (EOS) and for the
search of a predicted transition to the a phase of matter, quark-gluon
plasma (QGP). At present the transverse flow of particles is believed
to be one of the most clear signals to detect the creation of the QGP
in heavy-ion experiments (for recent review, see \cite{QM97,QM99}).
This explains the great interest of both experimentalists and 
theoreticians in the transverse flow phenomenon (see, e.g., 
\cite{StGr86,ReRi97,Oll98} and references therein), which was
predicted about 25 years ago \cite{SMG74} in nuclear shock wave
model analysis.

Initially, the collective flow has been conventionally subdivided into 
the radial flow, which is azimuthally symmetric, the bounce-off or 
directed flow \cite{DaOd85} in the reaction plane along the impact 
parameter axis ($x$-axis), and the squeeze-out flow developing out of 
the reaction plane. 
The latter two components represent the anisotropic
part of the transverse flow and appear only in noncentral heavy-ion
collisions. The first observation of the transverse flow was made
by the Plastic Ball \cite{Gu84} and the Streamer Chamber \cite{Re84}
collaborations at the BEVALAC energies ($E_{lab} = 100${\it A\/} MeV 
- 1.8{\it A\/} GeV). Later on the directed flow of charged particles
has been detected by E877 collaboration at the AGS energies
($E_{lab} = 10.7${\it A\/} GeV) and by NA49 \cite{na49prl98} and WA98 
\cite{wa98} collaborations at the SPS energies ($E_{lab} =  158$
{\it A\/} GeV). 

The collective flow is a very suitable observable to characterise the 
reaction dynamics because it is extremely sensitive to the 
interactions between the particles. At intermediate (SIS) energies the
evolution of flow is mainly governed by the density and momentum
dependence of the long-range attractive and short-range repulsive 
nuclear forces in the medium, i.e., the nuclear mean field
\cite{Aich91,Fuchs96,Fopi}. With rising energy (AGS, SPS) the mean
field gets less important while new degrees of freedom, strings, come
into play. It has been shown also that the transverse flow
could carry the primary information about the softening of the EOS
due to the QGP creation \cite{Amprl91,Brprc94,Br99}, including the 
subsequent hadronization, as well as the relaxation of the excited 
matter to (local) thermal equilibrium. 

The advanced technique for the analysis of the flow at
high energies, based on the Fourier expansion of the particle 
azimuthal distribution, has been developed in Refs.
\cite{VoZh96,Vol97,PoVo98}.
The distribution of the particles in the azimuthal plane can be 
presented as 
\beq
\ds
\frac{d N}{d \phi} = a_0 \left[ 1 + 2 \sum_{n=1}^{\infty}
v_n cos(n\phi) \right] ,
\label{eq1}
\eeq
where $\phi$ is the azimuthal angle between the momentum of the 
particle and the reaction plane. The first two coefficients, $v_1$ 
and $v_2$, are the amplitudes of the first and second harmonics in the 
Fourier expansion of the azimuthal distribution, respectively. The 
asymmetric fraction of the collective flow is decomposed in this 
analysis into the directed (bounce-off) flow of particles emitted 
preferentially along the $x$-axis, and the elliptic component, which 
is developed mostly either along the $x$-axis or in the squeeze-out 
direction. The coefficient $v_2$, therefore, characterises the 
eccentricity of the flow ellipsoid \cite{VoZh96}.

The importance of the elliptic flow to study 
collective effects in heavy-ion collisions was first stressed in 
\cite{Oll92}. In this paper the rotation of the elliptic flow from
the squeeze-out direction to the bounce-off direction with rising
projectile energy was discussed as well. The alignment of the 
elliptic flow in the plane of the directed flow has been 
experimentally detected in Au+Au collisions at the AGS energies
\cite{E877prl94} and in Pb+Pb collisions at the SPS energies
\cite{na49prl98,wa98npa98}. 
In \cite{Sorplb97} the sensitivity of the elliptic flow to the early
pressure was noticed. The elliptic flow seems to be generated only
during the very beginning of the collective expansion 
\cite{Sorplb97,KSHprl99}, while the radial flow is developing almost
until the freeze-out.
It was also pointed out that the elliptic flow should have a 
kinky structure \cite{Sor99} if the expanding and cooling fireball
undergoes a first-order phase transition from the QGP to hadrons.
Therefore, the characteristic features of the plasma hadronization can 
be traced by the dependence of the elliptic flow on the impact
parameter \cite{HeLe99}.

Although the collective flow is a unique complex phenomenon, the 
variety of its signals is very rich. Lacking a first principles 
theoretical description of heavy-ion collisions, one definitely
needs to explore semi-phenomenological models whose numerical
predictions can be compared with the experimental data on nuclear
collective effects in a wide energy range. These models can be 
classified in general either as macroscopic models or as microscopic 
ones. Macroscopic models are based on the hypothesis of (local)
thermal equilibrium in the system achieved by the large number of
various inelastic and elastic processes in the course of a nuclear
collision. The many-body distribution functions, which characterise
the nonequilibrium states, are rapidly reduced to the one-particle
distribution functions (one for each particle species), and the
kinetic stage emerges. At a longer time scale the system can reach
the hydrodynamic stage, where its evolution is described in terms
of the moments of the one-particle distribution functions, such as
average velocities, energies, and number of particles.
The evolution of a relativistic perfect fluid obeys the conservation
of energy and momentum \cite{LaLiv6}
\beq
\ds
\partial _{\mu} T^{\mu \nu} = 0 ,
\label{eq2}
\eeq
where
\beq
\ds
T^{\mu \nu} = (\varepsilon + P) u^\mu u^\nu + P g^{\mu \nu}
\label{eq3}
\eeq
is the energy-momentum tensor, and $\varepsilon,\ P,\ u^\mu$ are the
energy density, pressure, and local four-velocity, respectively.

Without the EOS, which links the pressure $P$ to the energy density
$\varepsilon$, the system of hydrodynamic equations (\ref{eq2}) - 
(\ref{eq3}) is incomplete.
Usually, the EOS is taken in a form
\beq
\ds
P = a\varepsilon \equiv c_s^2 \varepsilon ,
\label{eq4}
\eeq
with $c_s$ being the speed of sound in the medium. Inserting 
different equations of state, particularly with and without the
plasma EOS, into the one-, two-, or three-fluid hydrodynamic model 
\cite{GKLR86,ClSt86,MRSnpa89,Brnpa97} 
one can study the properties of the particle collective
flow at various incident energies \cite{Amprl91,Brprc94,Br99,Oll92}.

The microscopic models, developed to describe heavy-ion collisions 
in a wide range of bombarding energies, e.g.
\cite{lund,dpm,venus,qgsm,rqmd,hijing,vni,art,urqmd} and others,
do not rely on the assumption of thermal
equilibrium. They employ a dynamical picture of heavy-ion
interactions, in which the parton-, string-, and transport approaches
can be relevant. Though these models do not explicitly assume the 
formation of the QGP, the creation of the field of strongly interacted 
coloured strings may be considered as a precursor of the quark-gluon
plasma. Because of the uncertainties in the description of the early 
stage of heavy-ion collisions at ultrarelativistic energies, the 
microscopic and macroscopic models can be merged to implement the
phase transition to the deconfined phase directly in the microscopic
model \cite{Sor99,Baprc1,Baprc2}.  

In the present paper two microscopic models, QMD and QGSM, are 
employed to study the anisotropic flow components in collisions of
light and heavy ions at energies from SIS to SPS. The main goal is
to understand to what extent the characteristic signals of the hot
nuclear matter can be reproduced without invoking the assumption of
QGP creation. In other words, if the experimental data will
noticeably diverge from the results of simulations, this can be 
considered as an indication on new processes not included into
the models. The paper is organised as follows. A
brief description of the models is given in Sec.~\ref{sec2}.
Sections~\ref{sec3} and \ref{sec4} present the mass and impact 
parameter dependence of the simulated directed and elliptic flow, 
$v_1$ and $v_2$, at SIS, AGS, and SPS energies. 
Results obtained are discussed in Sec.~\ref{sec5}.
Finally, conclusions are drawn in Sec.~\ref{sec6}.

\section{Models}
\label{sec2}

The dynamics of nucleus-nucleus collisions at energies up to 
$\sqrt{s} \approx 2${\it A} GeV per nucleon can be described in terms
of reactions between hadrons and their excited states, resonances.
At higher energies additional degrees of freedom, i.e. strings, should 
be taken into account to describe correctly the processes of 
multiparticle production. Therefore, we employ the quantum molecular
dynamics (QMD) model \cite{Aich91,Bat94} at the SIS energies, while 
at the AGS and SPS energies the quark-gluon string model (QGSM) 
\cite{qgsm} is applied.

In the QMD approach the particles are propagated according to 
Hamilton equations of motion until their mutual interactions.
Each nucleon is represented by a Gaussian-shaped density in the
phase space. The black disk approximation is used to determine
the binary collision of hadrons. It implies that two hadrons can
collide if the centroids of two Gaussians are closer than the
distance $d_0 = \sqrt{\sigma_{\rm tot}(\sqrt{s})/\pi}$ during their
propagation. 
The Pauli principle is taken into account by blocking the collision
if the final states are already occupied in the phase space by
other particles. Among the inelastic channels the $\Delta(1232)$ 
resonance is the dominant one. Pion production takes place via 
resonance decay of $\Delta (1232)$ and $N^\ast(1440)$. Pions, which 
can propagate freely, i.e. without any 
mean field, undergo, however, a complex chain of reabsorption and
subsequent resonance decay processes before their freeze-out
\cite{uma97}. At SIS energies the reaction dynamics is governed
by the interplay between the nuclear mean field and binary 
collisions collisions, which pay a minor role at low energies
$(~ 100${\it A} MeV) due to the Pauli-blocking of possible 
scattering states, but become more and more important with rising
incident energy. In the present work we employ Skyrme-type 
mean field with a density dependence corresponding to a hard
EOS ($K=380$ MeV) and  momentum dependence fitted to the empirical
nucleon-nucleus optical potential \cite{Aich91}. This type of 
interaction has been shown to give a good description of flow data 
in the considered energy range around 1{\it A} GeV \cite{Fopi}.
Note, that the in-medium cross-section as well as the mean field can 
also be based on microscopic many-body approaches like Br\"uckner
theory \cite{Fuchs96,Bat94,Tue92,Boel99,Gait99} which is, however,
not the scope of the present investigation.
This approach
allows to describe heavy-ion collisions at energies up to few
{\it A} GeV. At higher energies the strings come into play.

The QGSM is based on the $1/N_c$ (where $N_c$ is the number of quark 
colours or flavours) topological expansion of the amplitude for 
processes in quantum chromodynamics and string phenomenology of 
particle production in inelastic binary collisions of hadrons. The 
diagrams of various topology, which arose due to the $1/N_c$ 
expansion, correspond at high energies to processes with exchange of 
Regge singularities in the $t$-channel. For instance, planar and 
cylindrical diagram corresponds to the Reggeon and Pomeron exchange, 
respectively.  Therefore, QGSM treats the elementary hadronic 
interactions on the basis of the Gribov-Regge theory, similar to the 
dual parton model \cite{dpm} and the VENUS model \cite{venus}. The 
model simplifies the nuclear effects and concentrates on hadron 
rescattering. As independent degrees of freedom QGSM includes octet 
and nonet vector and pseudoscalar mesons, and octet and decuplet 
baryons, and their antiparticles. 

The formation of the quark-gluon plasma is not assumed in the present 
version of the model. Thus, the effects similar to softening of the 
EOS in ultrarelativistic heavy-ion collisions, discussed below, are 
merely attributed to the dynamics of hadron rescattering and nuclear 
shadowing. We start from the study of energy and centrality dependence 
of the directed flow. 
 
\section{Directed flow}
\label{sec3}

For the simulations at all three energies, namely 1{\it A} GeV,
11.6{\it A} GeV, and 160{\it A} GeV, we choose light 
($^{32}$S+$^{32}$S) and heavy ($^{197}$Au+$^{197}$Au and 
$^{208}$Pb+$^{208}$Pb) symmetric systems. The directed and
elliptic flow of nucleons and pions as a function of rapidity
is defined as 
\beq
\ds
v_n^i = \cos{(n\phi_i)} \equiv \cos{(n\phi_i)}
\frac{d{\cal N}^i}{d y} \, dy \left/ \frac{d{\cal N}^i}{d y}
\, dy \right. ,
\label{eq5}
\eeq
where $n = 1,2$ and $i = N, \pi$. The mean directed and elliptic flow 
integrated over the whole rapidity range is simply
\beq
\ds
\langle v_n^i \rangle = \langle \cos{(n\phi_i)} \rangle \equiv 
\int \cos{(n\phi_i)}
\frac{d{\cal N}^i}{d y} \, dy \left/ \int \frac{d{\cal N}^i}{d y}
\, dy \right. ,
\label{eq6}
\eeq
To compare different systems
colliding at different energies the reduced rapidity $\tilde{y} = 
y/y_{proj}$ and reduced impact parameter $\tilde{b} = b/b_{max}$
has been used. The maximum impact parameter for a symmetric system is 
$b_{max} = 2\, R_A$.  The value of $\tilde{b}$ in the simulations is
varying from 0.15 (central collisions) up to 0.9 (most peripheral
collisions).
 
\begin{figure}[htp]
\centerline{\epsfysize=18cm \epsfbox{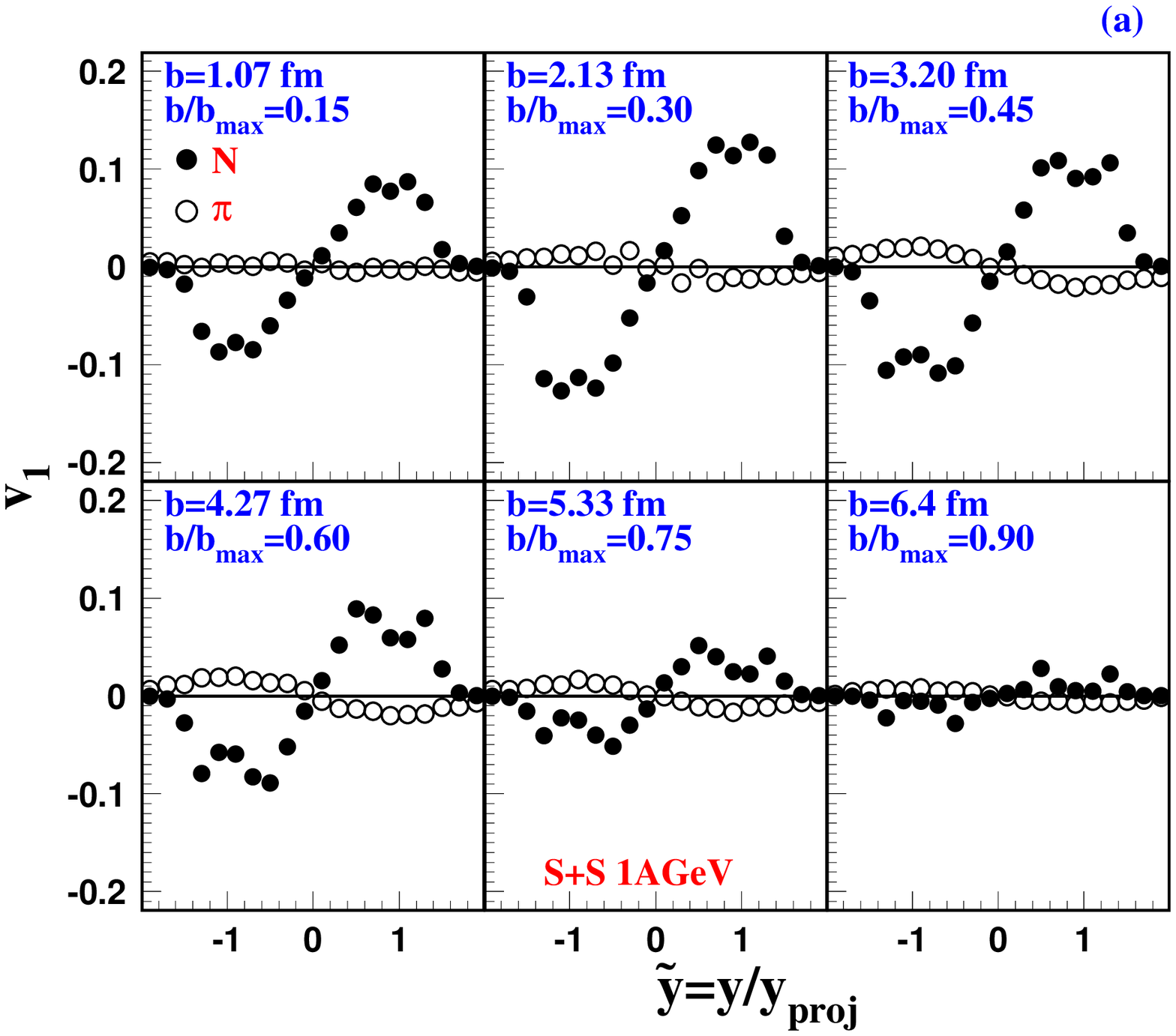}}
\caption{(a) Directed flow of nucleons (full circles) and pions 
(open circles) as a function of rapidity in $^{32}$S+$^{32}$S 
collisions at 1{\it A} GeV.\\
(b) the same as (a) but for $^{197}$Au+$^{197}$Au collisions.
}
\centerline{\epsfysize=18cm \epsfbox{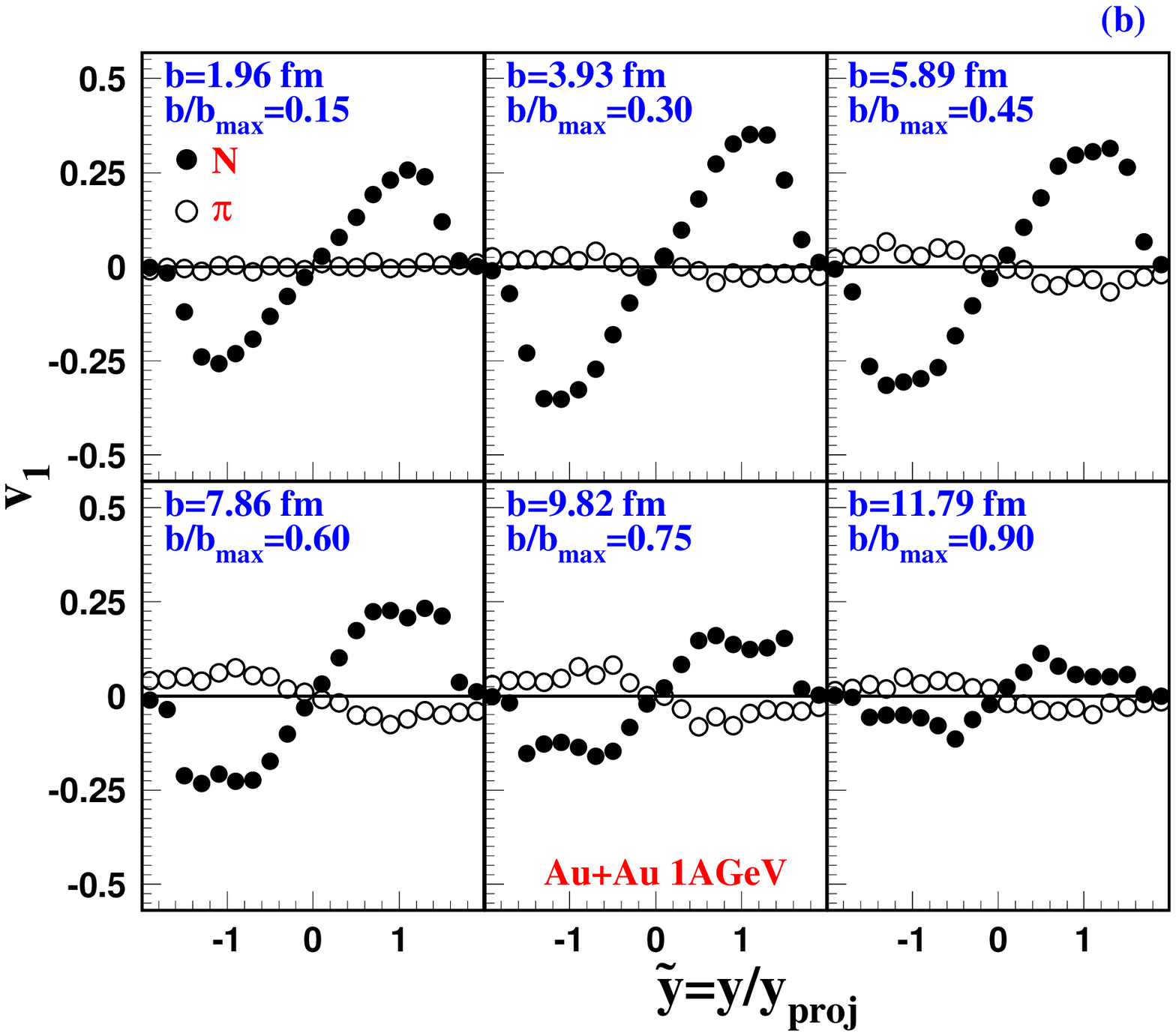}}
\label{fig1}
\end{figure}

The rapidity distributions of $v_1$ at SIS energies are shown in 
Fig.~\ref{fig1}(a) for S+S and in Fig.~\ref{fig1}(b) for Au+Au
system. The directed flow of nucleons has a characteristic
$S$-shape attributed to the standard $\langle p_x/A \rangle$
distribution. Conventionally, we will call this type of flow,
for which the slope $d v_1/d \tilde{y}$ is positive, {\it normal\/} 
flow, in contrast to the {\it antiflow\/} for 
which $d v_1/d \tilde{y} < 0$ in the midrapidity region.

The nucleon flow reaches the maximum at $\tilde{b} = 0.3-0.45$
both in S+S and Au+Au system, and then it drops. In the 
midrapidity range the flow can be well approximated by a linear 
dependence. The slope parameters of the $v_1^N (y)$ distributions are 
listed in Table~\ref{tab1} together with the $d v_1^N/d \tilde{y}$
data at higher energies. Pions at SIS energies show only weak
antiflow which reaches a maximum around $\tilde{b} = 0.45-0.6$ for 
S+S and $\tilde{b} = 0.6-0.75$ for Au+Au collisions, i.e. in more 
peripheral collisions compared to the maximal nucleon directed flow.
This behaviour is understandable since the evolution of a positive
nuclear flow due to the (momentum dependent) repulsive $NN$-forces
requires sufficiently large participant matter, whereas the negative 
pion flow due to shadowing needs large spectators. 
The antiflow of pions can also be fitted by a linear dependence;
slope parameters are presented in Table~\ref{tab1}.

\begin{figure}[htp]
\centerline{\epsfysize=18cm \epsfbox{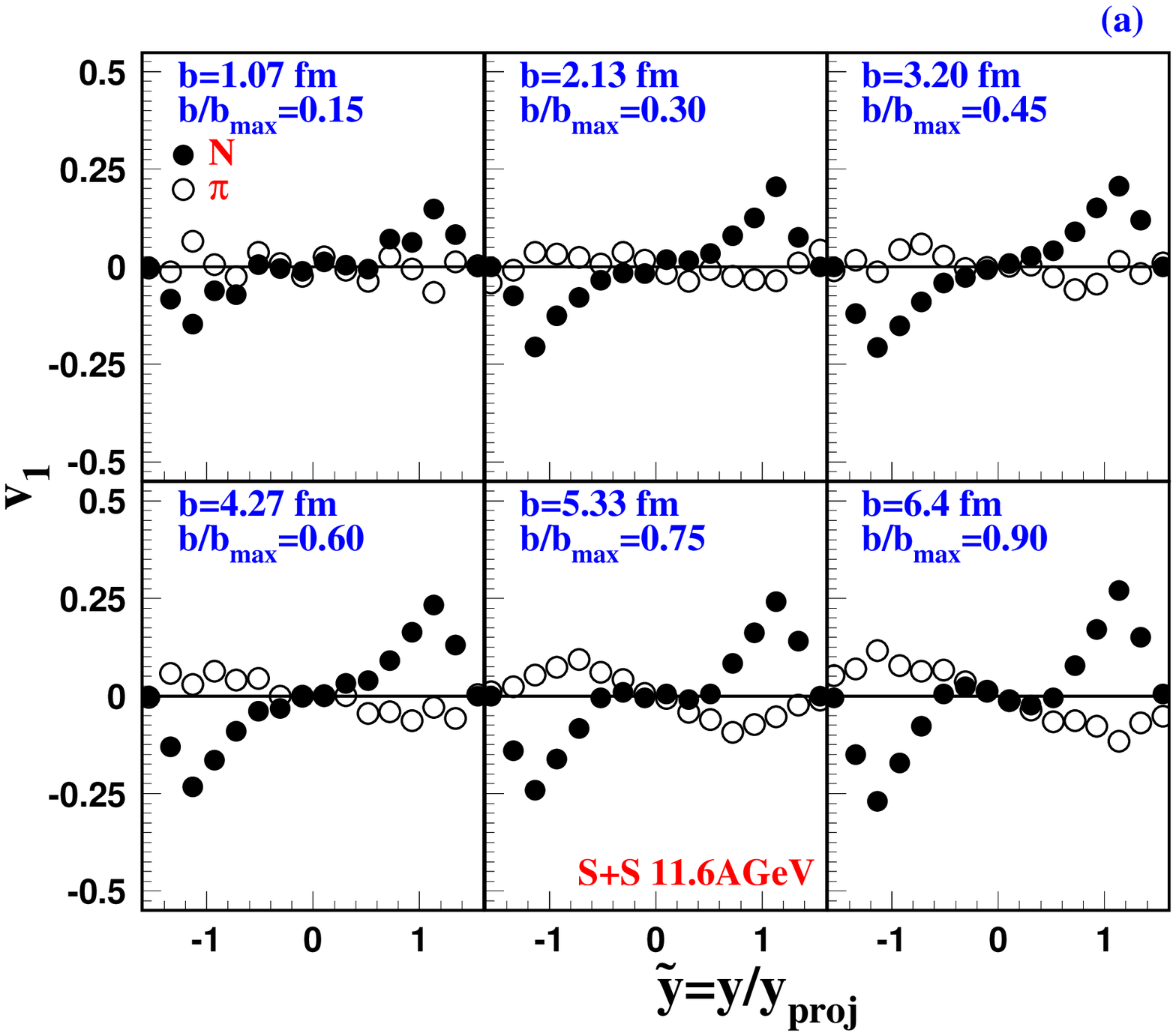}}
\caption{(a) Directed flow of nucleons (full circles) and pions 
(open circles) as a function of rapidity in $^{32}$S+$^{32}$S 
collisions at 11.6{\it A} GeV.\\
(b) the same as (a) but for $^{197}$Au+$^{197}$Au collisions.
}
\centerline{\epsfysize=18cm \epsfbox{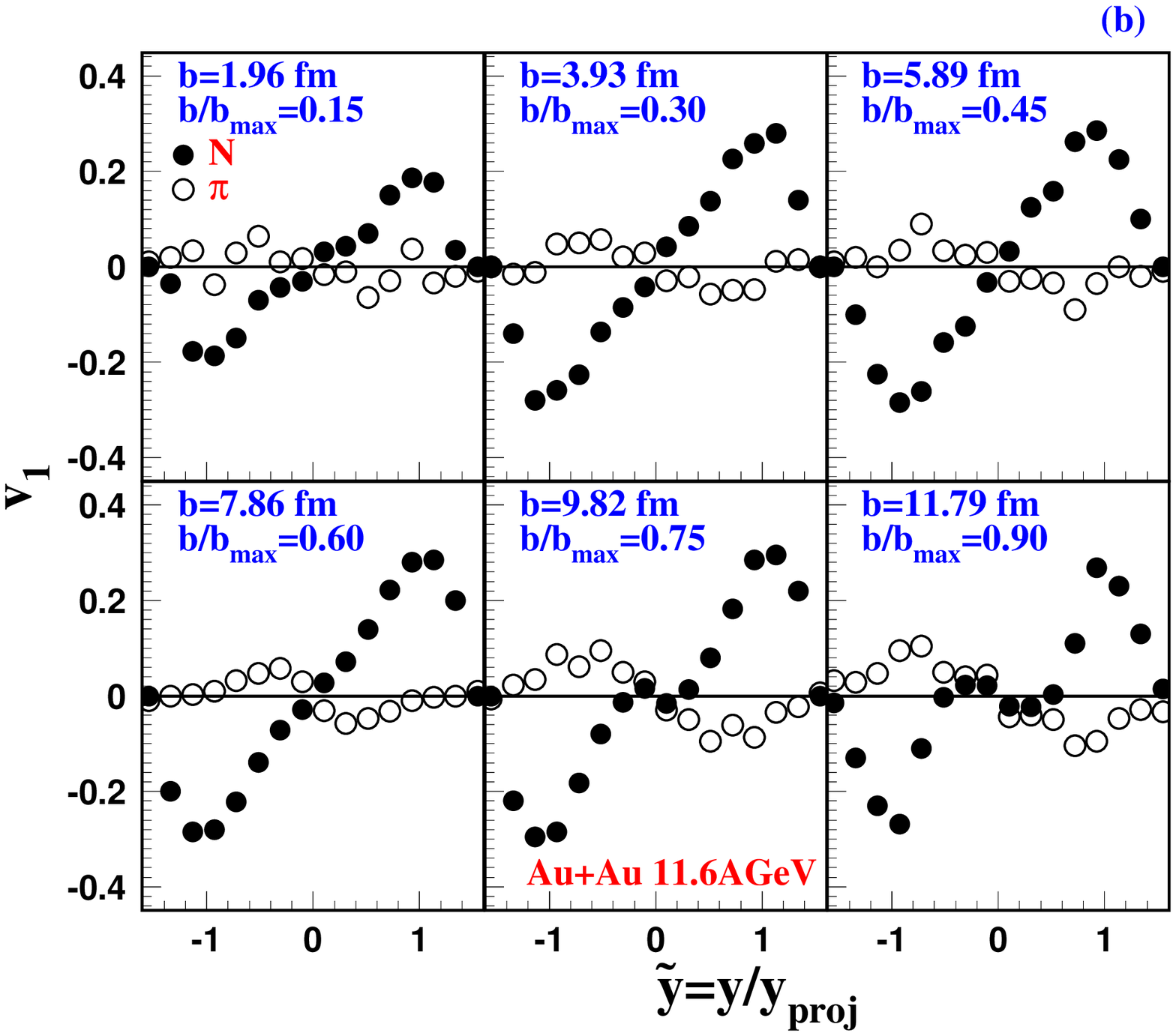}}
\label{fig2}
\end{figure}

The directed flow $v_1^i(y)$ calculated for the same systems, S+S
and Au+Au, at AGS energies is shown in Fig.~\ref{fig2}(a) and
Fig.~\ref{fig2}(b), respectively. Here the distributions for 
nucleons differ considerably, especially in light system, from
those at 1{\it A} GeV. The deviations of $v_1^N(y)$ from the 
straight line in S+S collisions begin noticeable already at 
$\tilde{b} = 0.3$. The nucleon directed flow goes to zero in the 
midrapidity range with increasing impact parameter. Moreover, 
even antiflow is developed in very peripheral collisions at 
$\tilde{b} = 0.9$, as seen in Fig.~\ref{fig2}(a).

In contrast, in heavy Au+Au collisions at 11.6{\it A} GeV there are 
no singularities in the behaviour of $v_1^N(y)$ up to $\tilde{b}=0.6$. 
The plateau in the midrapidity region seems to 
build up only at $\tilde{b} \geq 0.75$, see Fig.~\ref{fig2}(b).
Pion directed flow has negative slope in the midrapidity range for 
both light and heavy colliding system. Values of the slope parameter
are listed in Table~\ref{tab1}.
It is worth to mention that $v_1^{\pi}$ increases as the reaction
becomes more peripheral, and that both pion and nucleon directed flow
does not vanish even at $\tilde{b} = 0.9$ compared to the flow at
SIS energies.

\begin{figure}[htp]
\centerline{\epsfysize=18cm \epsfbox{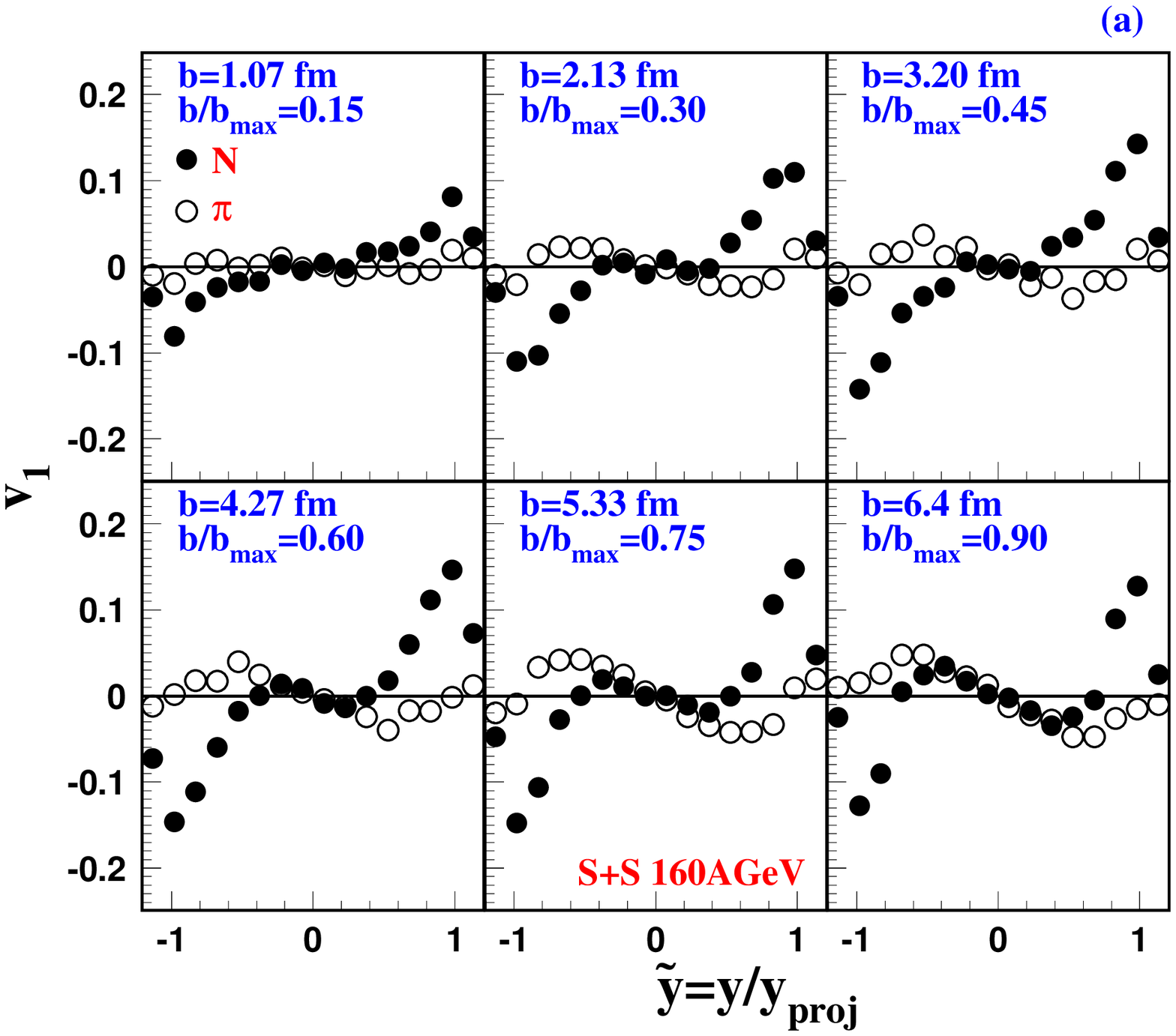}}
\caption{(a) Directed flow of nucleons (full circles) and pions 
(open circles) as a function of rapidity in $^{32}$S+$^{32}$S 
collisions at 160{\it A} GeV.\\
(b) the same as (a) but for $^{208}$Pb+$^{208}$Pb collisions.
}
\centerline{\epsfysize=18cm \epsfbox{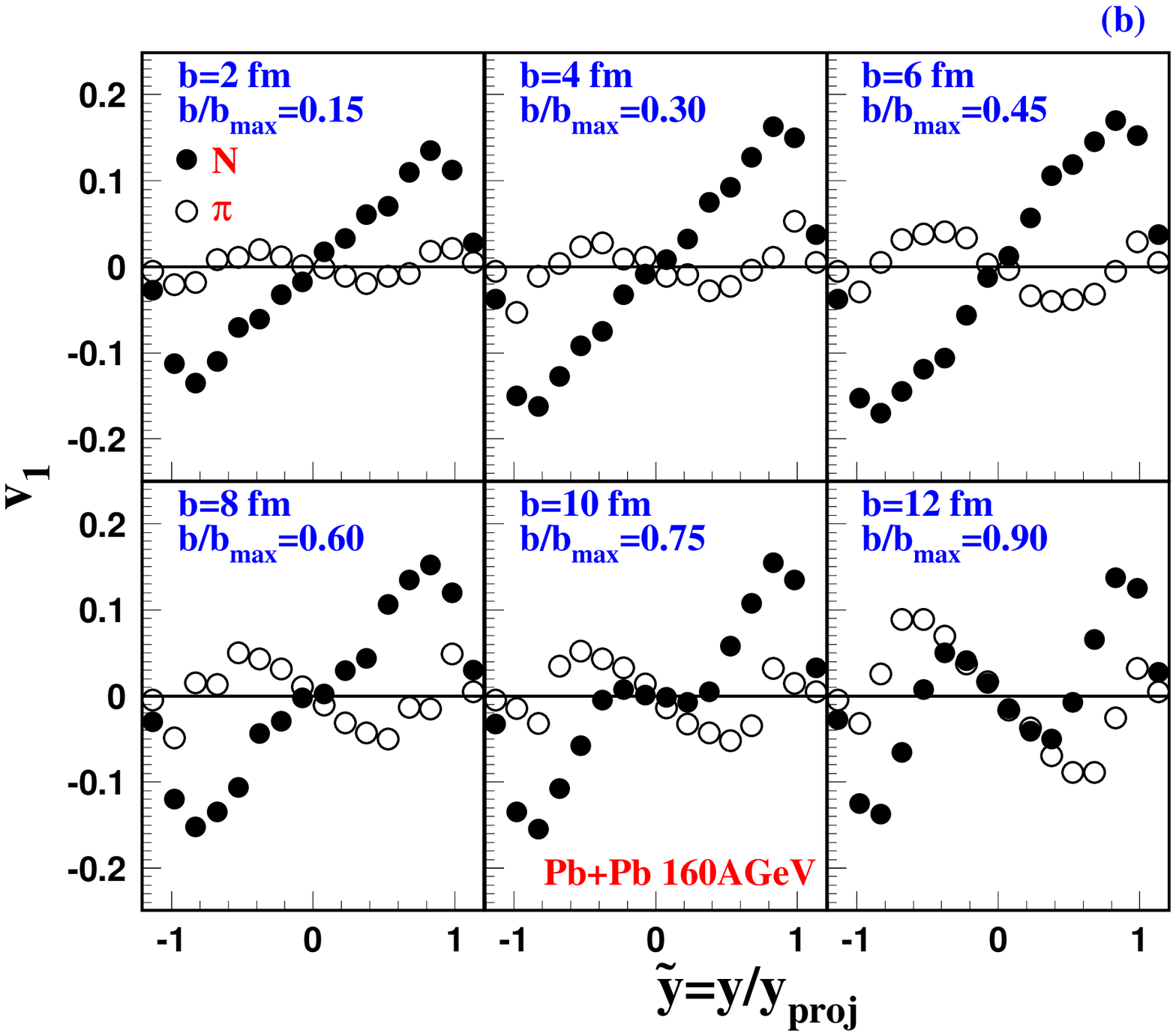}}
\label{fig3}
\end{figure}

At the SPS energies the directed flow of nucleons has negative slope 
in the midrapidity region already in semiperipheral S+S collisions
as depicted in Fig.~\ref{fig3}(a). With the increase of the impact 
parameter the nucleon antiflow becomes stronger. The $y$-dependence 
of $v_1^N$ in Pb+Pb collisions is shown in Fig.~\ref{fig3}(b).
Here the deviations from the straight line start to develop at 
$\tilde{b} = 0.45$ in the central rapidity window, 
$|\tilde{y}| \leq 0.25$. It is interesting that the slope of the
antiflow of nucleons at $\tilde{b} = 0.9$ is similar to that of the 
pion flow. The latter reaches maximum also at $\tilde{b} = 0.9$
in Pb+Pb, as well as in S+S collisions.

The disappearance of the directed flow of hadrons can be regarded as 
an indication for a softening of the EOS \cite{HuSh95,RiGy96} due to
a QGP-hadron phase transition. The simple hypothesis would be
that, despite the absence of the plasma formation in the microscopic 
model, the colour field of quark-antiquark and quark-diquark strings 
can force the softening of the hadronic EOS. This idea explains the 
disappearance of the directed flow at energies of AGS and higher, but 
obviously fails to explain why the effect is stronger in peripheral 
collisions and for light systems like S+S. The correct explanation can 
be, therefore, that the apparent softening of the equation of state is
in fact caused by the nuclear shadowing \cite{Br95,BZFF99}. The
mechanism of the development of nuclear antiflow in the 
midrapidity range of nuclear peripheral collisions is elaborated 
in Sec.~\ref{sec5}.

\begin{figure}[htp]
\centerline{\epsfysize=18cm \epsfbox{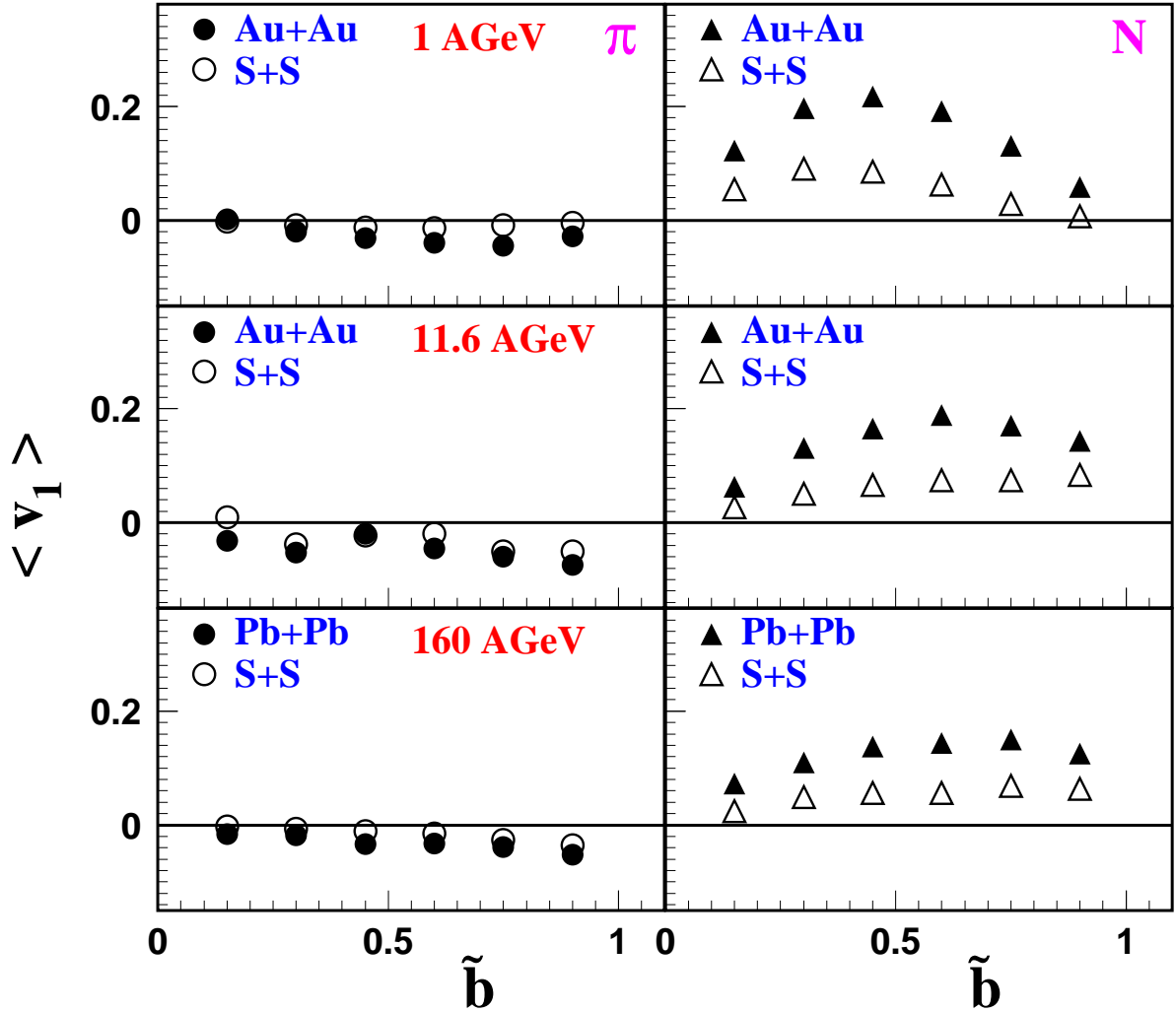}}
\caption{The mean directed flow of nucleons (full circles) and
pions (open circles) in light and heavy system colliding at 
1{\it A} GeV, 11.6{\it A} GeV, and 160{\it A} GeV, respectively.
}
\label{fig4}
\end{figure}

The mean directed flow $\langle v_1 \rangle$ of pions and nucleons is
shown in Fig.~\ref{fig4}. Except nearly central events, the pion
mean flow is negative for both light and heavy colliding system at
all three energies. At the AGS and SPS energies $\langle v_1^\pi
\rangle$ rises steadily as the reaction becomes more peripheral.
The mean directed flow of nucleons, which is always positive, seems
also to exhibit a similar tendency. The maximum in $\langle v_1^N (b)
\rangle$ distribution is located around $\tilde{b} = 0.4$ for Au+Au
collisions at 1{\it A} GeV. It is shifted to $\tilde{b} = 0.6$ at
11.6{\it A} GeV, and is completely dissolved at higher energies.

\section{Elliptic flow}
\label{sec4}

Since the elliptic flow develops at the very beginning stage of
nuclear collision, it might be even a better tool to probe the
nuclear EOS under extreme conditions \cite{Dan99}. Particularly,
calculations based on a relativistic hadron transport model 
indicate a transition of elliptic flow from out-of-plane to
in-plane for the case of the QGP formation in Au+Au collisions in 
the energy range $1 - 11${\it A} GeV \cite{Dan98}. Recent 
experimental data \cite{e895} confirm the transition from 
negative to positive elliptic flow at $E \approx 4${\it A} GeV,
which was considered as indication of the softening of
nuclear EOS. But can this change in the behaviour of elliptic flow
be induced by some other reasons? To answer the question the
microscopic study of elliptic flow of nucleons and pions has been
performed at energies from 1{\it A} GeV to 160{\it A} GeV. 
 
\begin{figure}[htp]
\centerline{\epsfysize=18cm \epsfbox{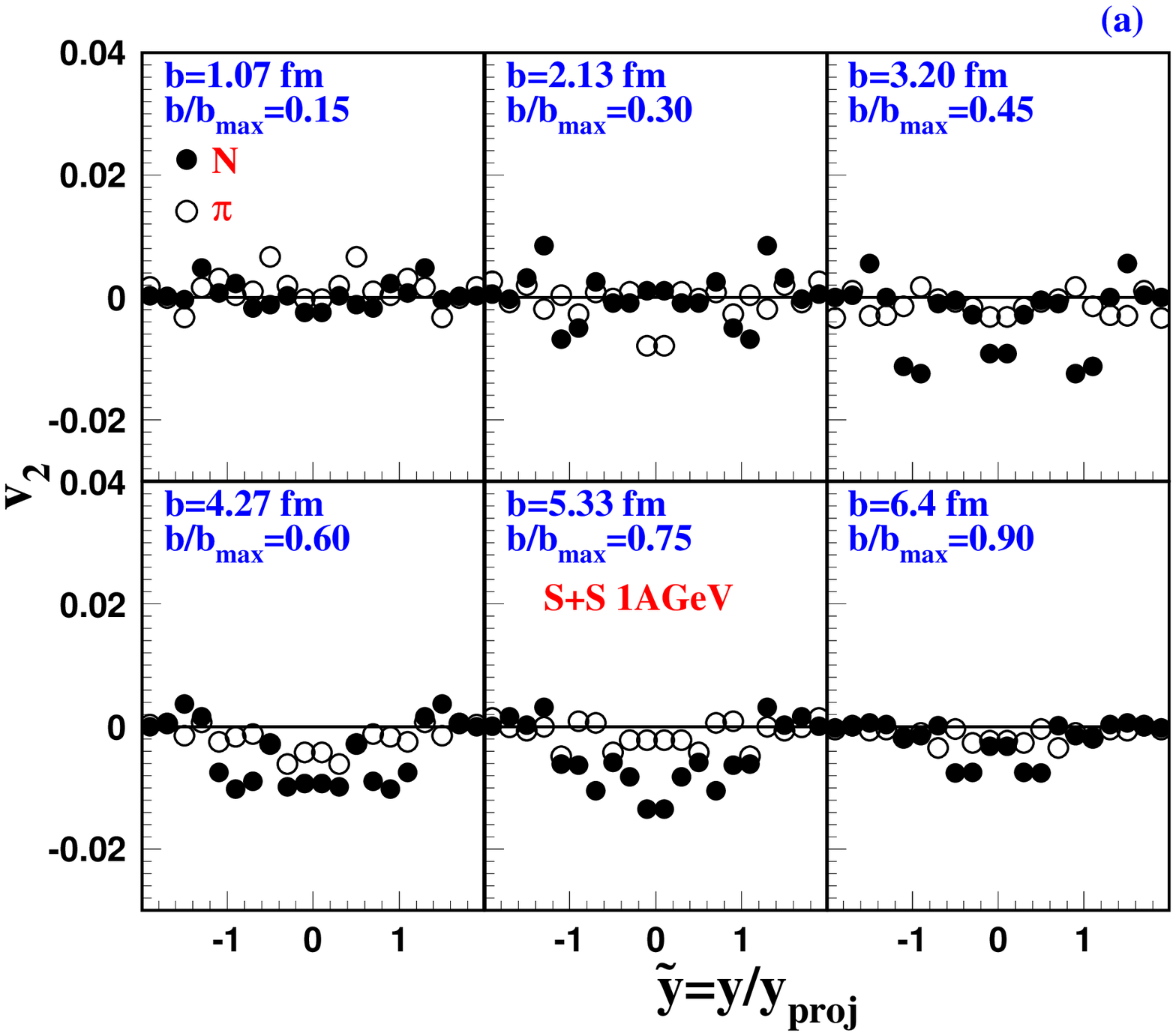}}
\caption{(a) Elliptic flow of nucleons (full circles) and pions 
(open circles) as a function of rapidity in $^{32}$S+$^{32}$S 
collisions at 1{\it A} GeV.\\
(b) the same as (a) but for $^{197}$Au+$^{197}$Au collisions.
}
\centerline{\epsfysize=18cm \epsfbox{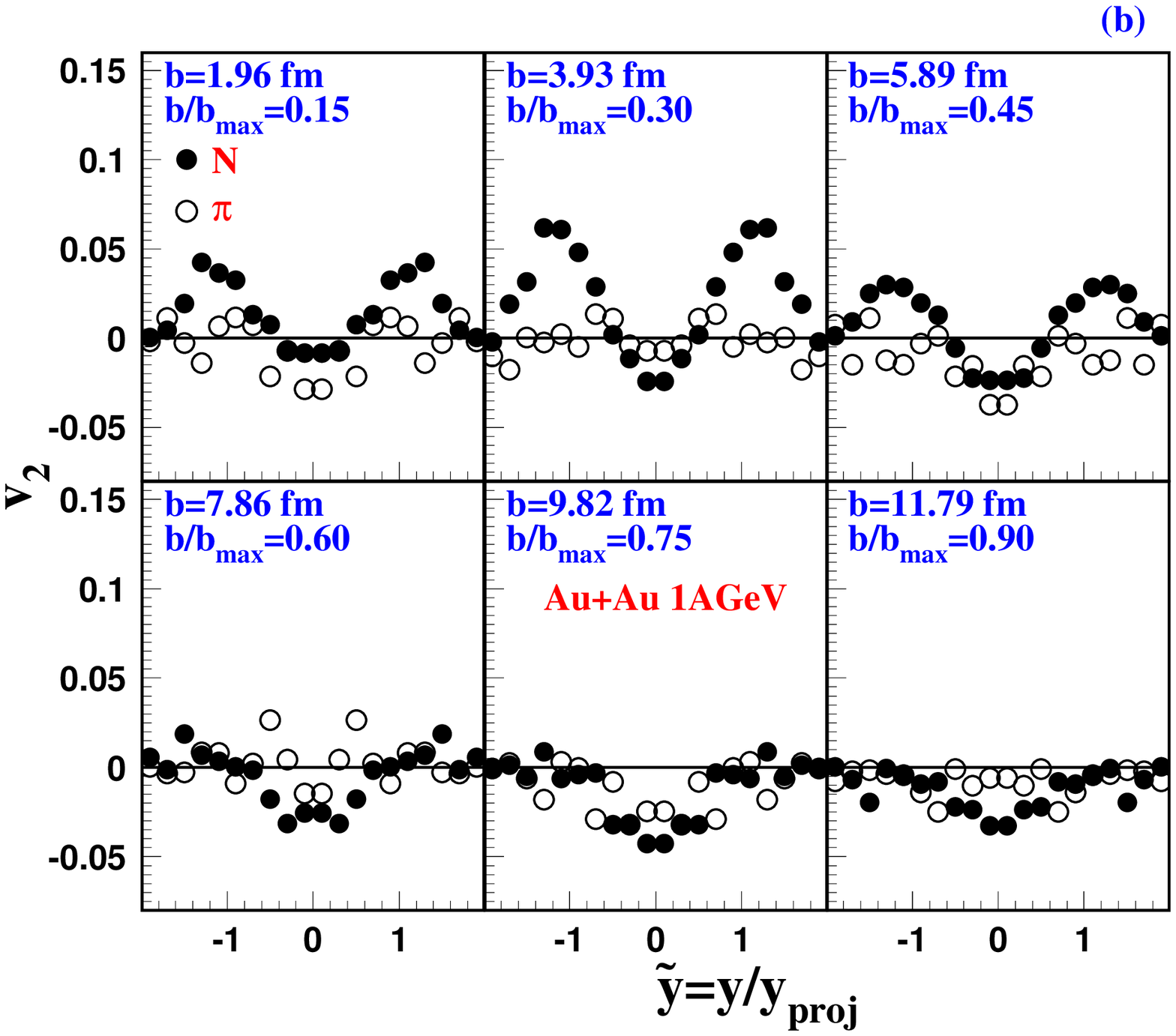}}
\label{fig5}
\end{figure}

Figures~\ref{fig5}(a) and \ref{fig5}(b) depict the elliptic flow of
pions and nucleons in S+S and Au+Au collisions, respectively, at 
1{\it A} GeV. The elliptic flow of pions is small and negative at
$\tilde{b} \geq 0.45$. Nucleon elliptic flow is also negative in
peripheral and semiperipheral collisions. The nucleon flow increases
to maximum at $\tilde{b}=0.75$ and then drops. For heavy system the 
pionic flow is negative already at $\tilde{b}=0.15$, while the 
nucleon flow at $\tilde{b} \leq 0.45$ has two positive peaks,
centred around $|\tilde{y}| \approx 1.4$, and the dip in the
midrapidity region, where $v_2^N$ is negative. In peripheral
collisions the positive peaks in the $v_2^N(y)$ distribution
vanish, and the negative elliptic flow of nucleons becomes
stronger. Generally, the spectator matter at target/projectile 
rapidities shows in-plane flow $(v_2^N > 0)$ whereas the
participant matter at midrapidity shows preferential out-of-plane
emission $(v_2^N < 0)$. 

\begin{figure}[htp]
\centerline{\epsfysize=18cm \epsfbox{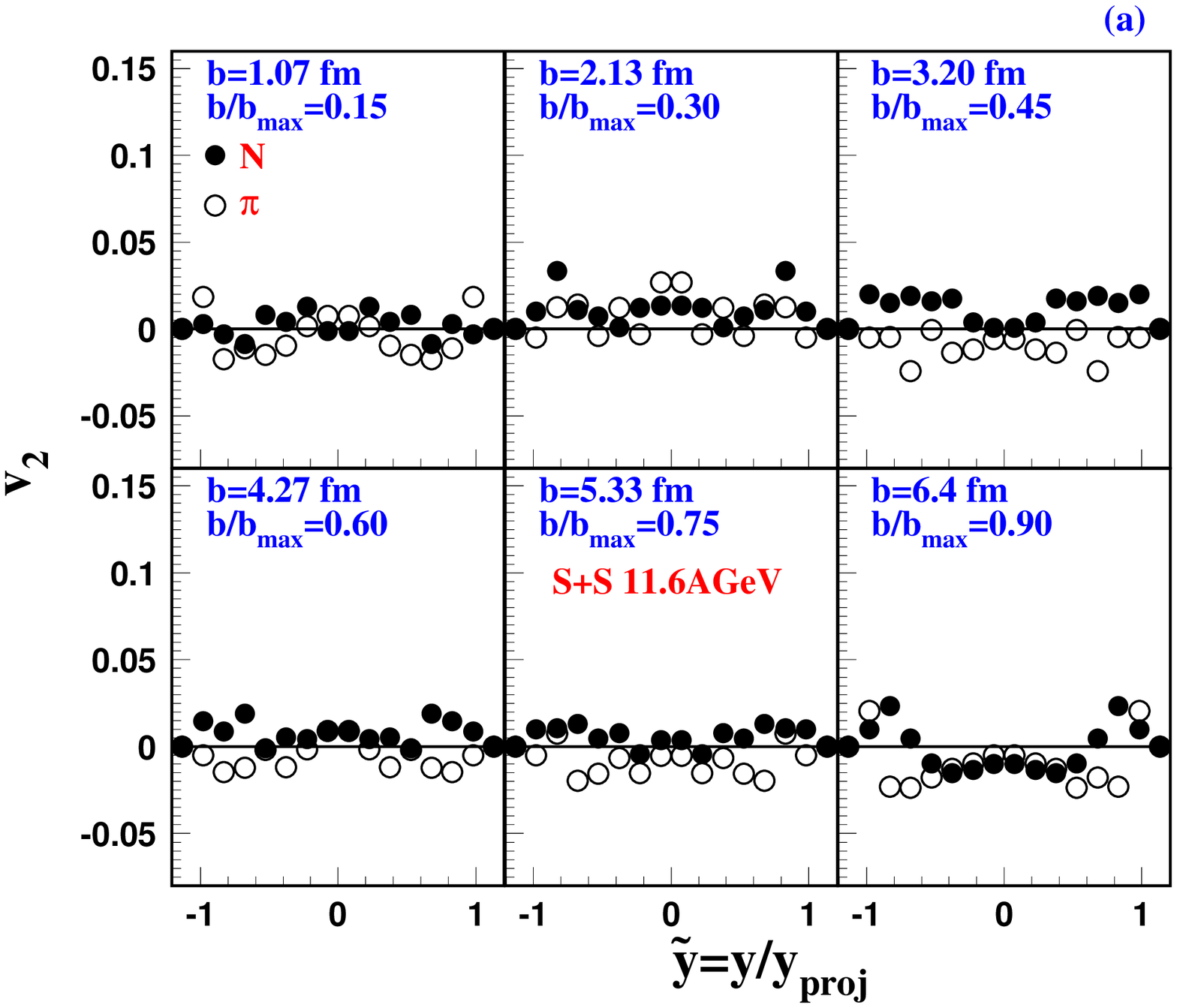}}
\caption{(a) Elliptic flow of nucleons (full circles) and pions 
(open circles) as a function of rapidity in $^{32}$S+$^{32}$S 
collisions at 11.6{\it A} GeV.\\
(b) the same as (a) but for $^{197}$Au+$^{197}$Au collisions.
}
\centerline{\epsfysize=18cm \epsfbox{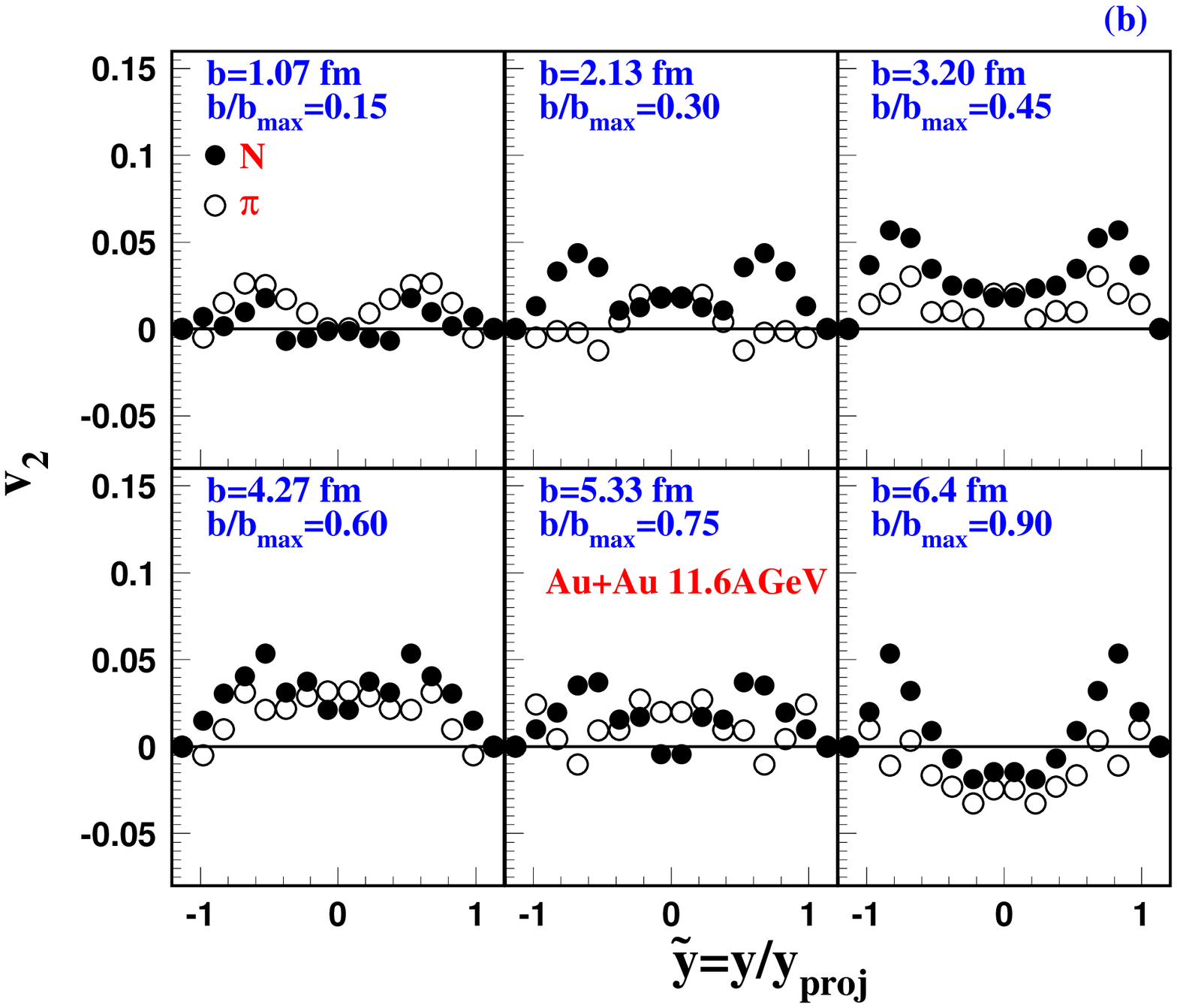}}
\label{fig6}
\end{figure}

The elliptic flow of nucleons at the AGS energies, shown in
Figs.~\ref{fig6}(a) and \ref{fig6}(b), also has a two-hump 
structure both in S+S and in Au+Au collisions. But in Au+Au 
interactions the nucleon flow becomes negative in the central part
of $\tilde{y}$-distribution only at $\tilde{b} \geq 0.75$. 
The peaks are quite noticeable and shifted closer to the 
center of the distributions. In Au+Au collisions the elliptic flow 
of both pions and nucleons is at least twice as large as in S+S
interactions. At $\tilde{b} \leq 0.6$ the nucleon flow is positive,
while the pionic flow becomes negative at midrapidity already at 
$\tilde{b} = 0.45$.

\begin{figure}[htp]
\centerline{\epsfysize=18cm \epsfbox{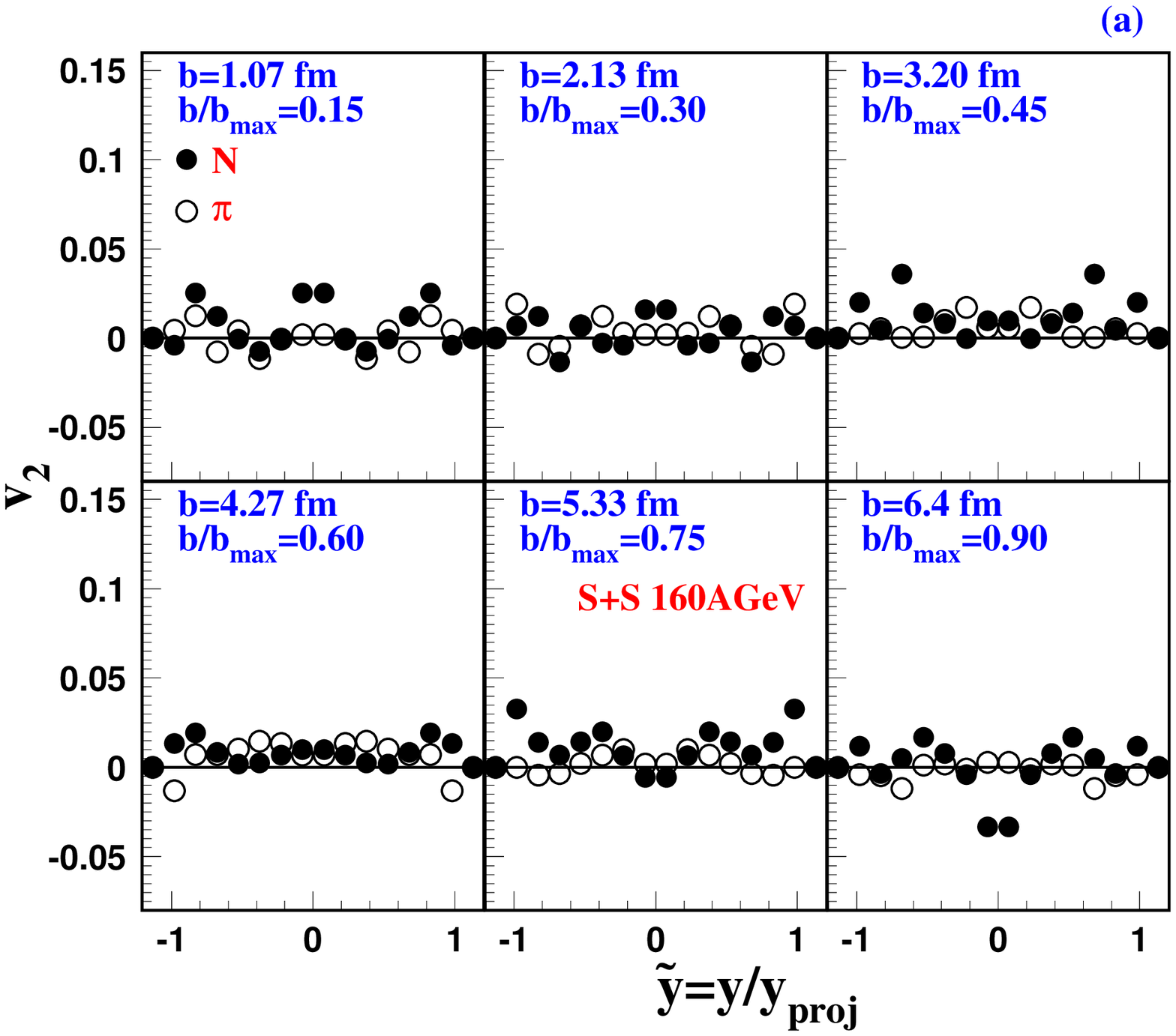}}
\caption{(a) Elliptic flow of nucleons (full circles) and pions 
(open circles) as a function of rapidity in $^{32}$S+$^{32}$S 
collisions at 160{\it A} GeV.\\
(b) the same as (a) but for $^{208}$Pb+$^{208}$Pb collisions.
}
\centerline{\epsfysize=18cm \epsfbox{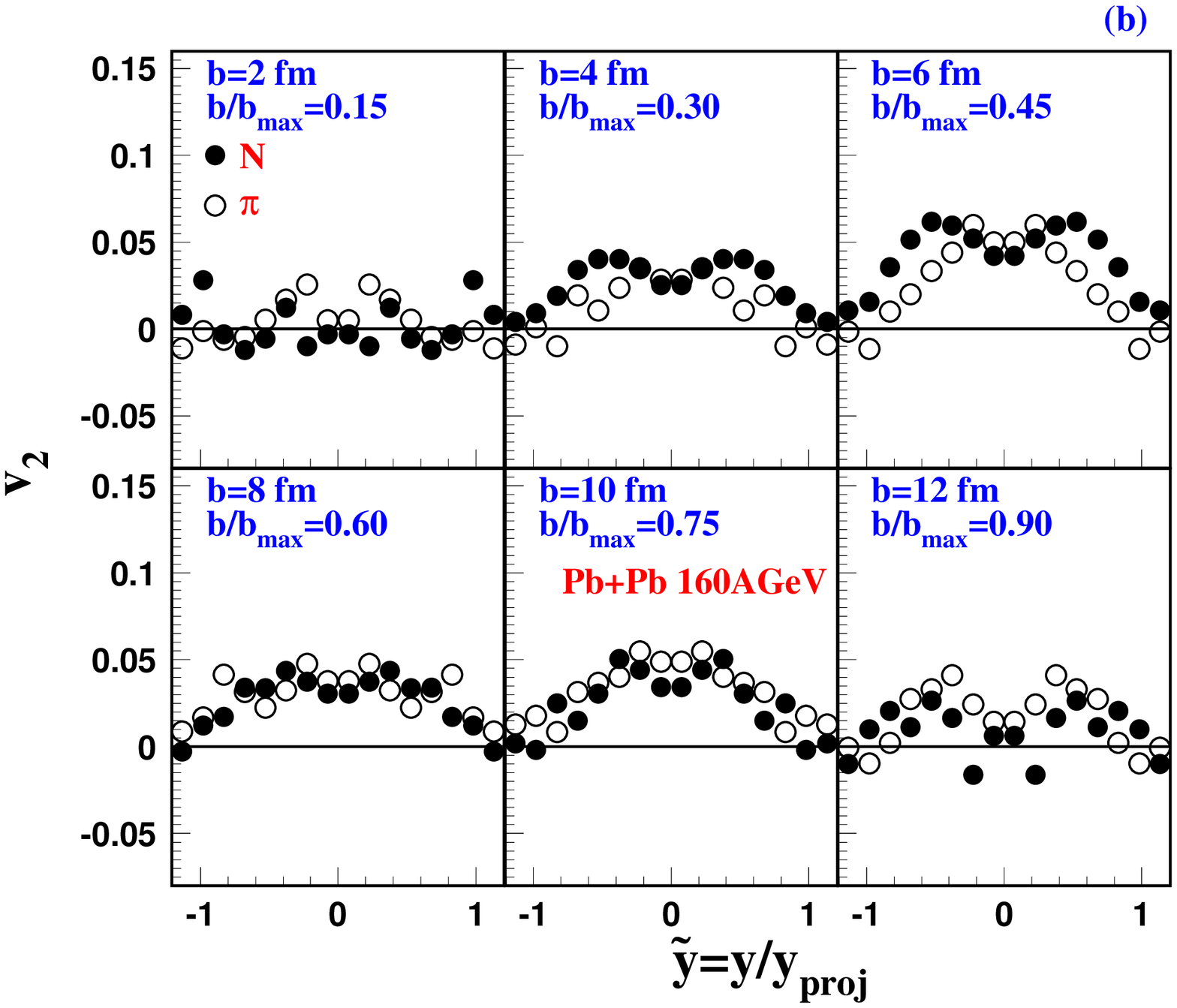}}
\label{fig7}
\end{figure}

At the SPS energies the elliptic flow of pions in S+S collisions
is quite flat and slightly positive, as demonstrated in 
Fig.~\ref{fig7}(a). The flow of nucleons in this reaction is also
small and positive except for $\tilde{b} \geq 0.75$, where the 
negative dip around $\tilde{y} = 0$ is built up. The negative flow
is seen for nucleons in Pb+Pb interactions only at $\tilde{b}=0.9$ 
in Fig.~\ref{fig7}(b). The origin of the negative values of $v_2^N$
at midrapidity can be linked to absorption of hadrons, emitted at
$\theta = 90^{\circ}$ angle in the reaction plane, by the dense
baryon rich spectators, while hadrons emitted out-of-plane remain
almost unaffected. Note also that the positions of positive
maxima in $v_2^N(y)$ distributions in Pb+Pb reactions are shifted
to $\tilde{y} \approx 0.45$. Compared to $v_2^\pi$ in S+S 
interactions, the elliptic flow of pions in Pb+Pb collisions is 
large and positive. It has a noticeable dip at midrapidity only in
very peripheral collisions. 

\begin{figure}[htp]
\centerline{\epsfysize=18cm \epsfbox{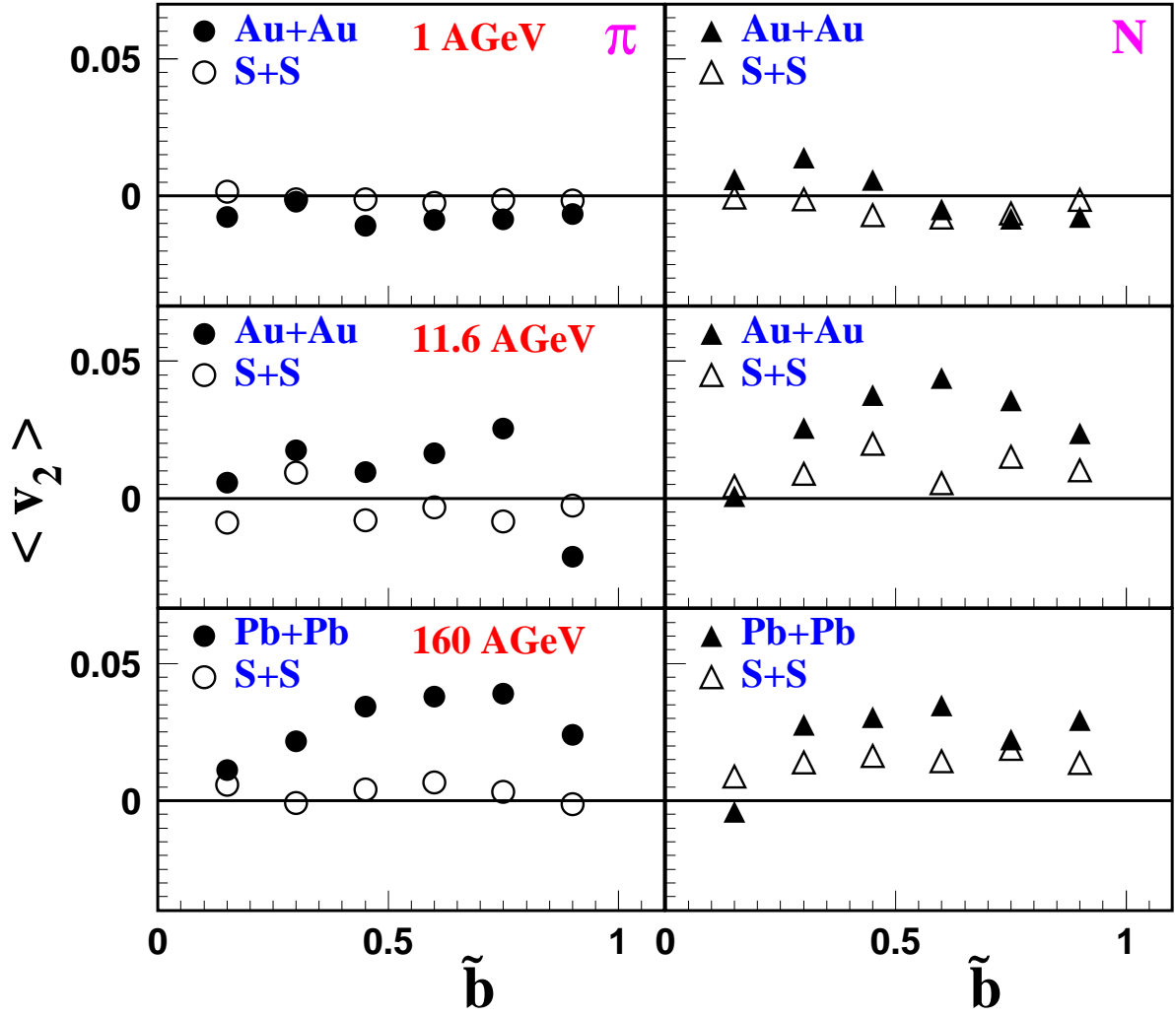}}
\caption{The mean directed flow of nucleons (triangles) and
pions (circles) in light (open symbols) and heavy (full symbols) 
system colliding at 
1{\it A} GeV, 11.6{\it A} GeV, and 160{\it A} GeV, respectively.
}
\label{fig8}
\end{figure}

The $\tilde{b}$-dependence of elliptic flow integrated over the 
whole rapidity range is presented in Fig.~\ref{fig8}. In S+S
collisions the mean elliptic flow of pions is quite weak for all
three energies, though it appears to change the sign from negative
at 1{\it A} GeV to positive at 160{\it A} GeV. The nucleon flow in
S+S reaction is more distinct. It is negative at the SIS energies,
while at both AGS and SPS energies $\langle v_2^N(\tilde{b})\rangle $
is positive and almost constant.

The situation is changed drastically with the rise of the mass 
number of colliding nuclei from $A = 32$ to $A = 197 (208)$.
The mean elliptic flow of pions becomes positive at the AGS energies,
in accord with the experimental results \cite{e895}. The flow reaches
maximum values at 160{\it A} GeV. It is easy to see that the strength
of $\langle v_2^\pi (\tilde{b}) \rangle$ increases with $\tilde{b}$
rising to 0.75, which corresponds to $b = 10$ fm in the calculations, 
and then drops. At the SIS energies the nucleon mean elliptic flow is 
positive in semicentral and semiperipheral events with $\tilde{b} 
\leq 0.45$ and negative at higher values of the impact parameter.
Thus in the heavy system there appears a 
transition of in-plane to out-of-plane flow with decreasing
centrality of the reaction. A detailed analysis of the EOS
dependence on the nuclear mean field at SIS energies will be 
presented elsewhere. 
The nucleon flow has the maximum strength at 11.6{\it A} GeV, in
contrast to the pion mean elliptic flow which rises continuously with
increasing incident energy.  
 
\section{Discussion of the results}
\label{sec5}

\begin{figure}[htp]
\centerline{\epsfysize=18cm \epsfbox{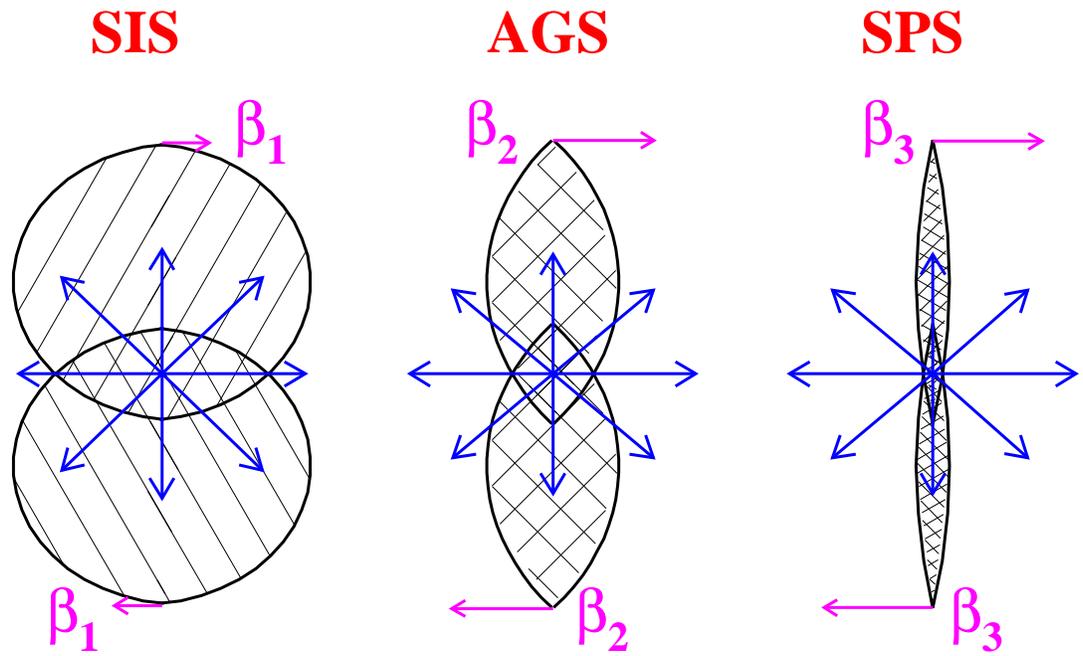}}
\caption{Symmetric system of colliding nuclei at maximum overlap
shown in (x,z)-plane at SIS, AGS, and SPS energies, respectively.
Arrows indicate the possible directions of particle emission from
the central zone.
}
\label{fig9}
\end{figure}

\begin{figure}[htp]
\centerline{\epsfysize=18cm \epsfbox{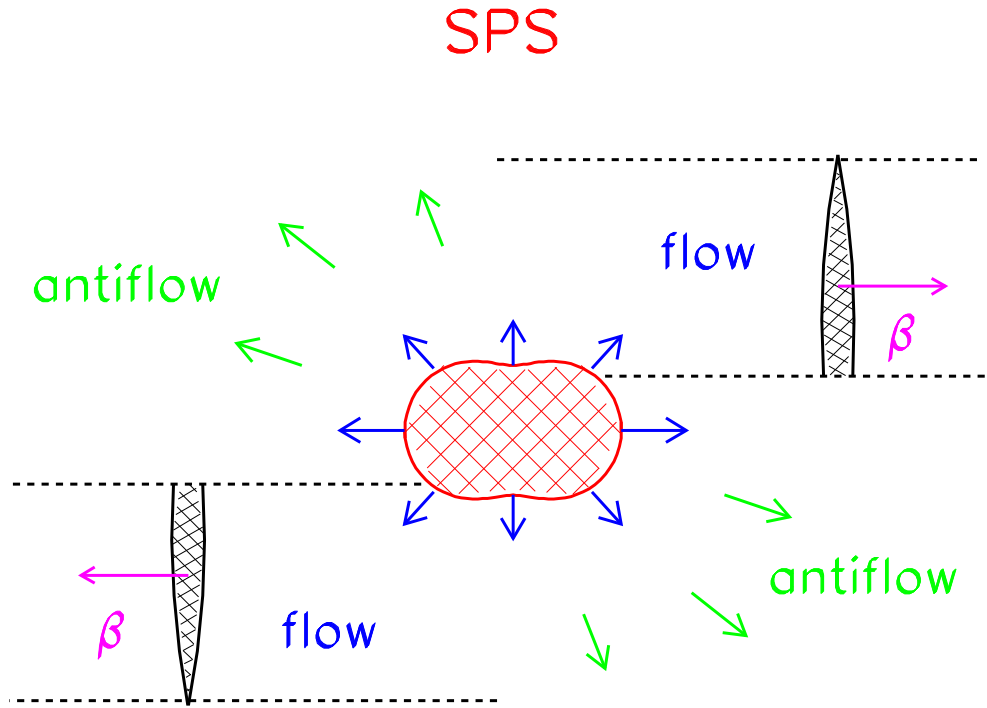}}
\caption{Formation of the antiflow in the midrapidity range in 
peripheral collisions at SPS energies. 
Particles emitted early in the ``normal" direction are absorbed
by the spectators, while particles emitted in the opposite direction 
(antiflow) remain unaffected. 
}
\label{fig10}
\end{figure}

We see in Sec.~\ref{sec3} that directed flow of nucleons in
(semi)peripheral heavy-ion collisions noticeably deviates from the
straight line behaviour in the midrapidity range at AGS energies or
higher. The comparison between light and heavy systems
colliding at the same energy per nucleon shows that the effect is 
stronger in the light system. The most probable explanation of this
phenomenon is nuclear shadowing. To clarify the idea the symmetric
system of two nuclei at maximum overlap is shown in Fig.~\ref{fig9}
for the three energies under consideration. In addition, 
Fig.~\ref{fig10} illustrates the development of the antiflow-like 
behaviour in the midrapidity region. As was discussed in, e.g.
\cite{Br95,BZFF99}, the total flow of hadrons is a result of
mutual cancellation of two competitive components, namely, the 
normal flow which follows the ongoing spectators, and the antiflow
which develops towards the baryon dilute areas of collision.
The normal flow integrated over the whole rapidity range is always
slightly larger than the integrated antiflow. But in the midrapidity
window the antiflow can dominate over its normal counterpart. For
instance, hadrons with small rapidities, emitted early in the 
direction of normal flow in heavy-ion collision at 160{\it A} GeV, 
(see Fig.~\ref{fig10}) will be absorbed by flying spectators. In 
contrast, hadrons, emitted in the direction of antiflow even at the 
angles close to $\theta = 90^\circ$ to the beam axis, propagate 
freely. 

This effect can be reduced by (i) increasing the centrality of the 
collision and (ii) by decreasing the center-of-mass energy of 
colliding nuclei (see Fig.~\ref{fig9}). In both cases the area where 
particles can be emitted without shadowing significantly shrinks. It 
is important to mention here that in heavy-ion collisions at collider
energies, RHIC ($\sqrt{s}=200$ GeV) and LHC ($\sqrt{s}=5.5$ TeV), 
the disappearance of nucleon directed flow in the midrapidity range
should emerge already in semicentral collisions with $b \leq 3$ fm.

But why the irregularities in $v_1^N(y)$-distribution start to 
develop in light system at smaller impact parameter compared to that
of heavy system? To answer this question note that the larger volume
of overlapping zone in heavy system leads to the intensive 
rescattering of baryons and increase of hadron emission from the 
central fireball. The spectators still absorb several early emitted
hadrons, but this process becomes less efficient compared to that of
the light system, where the isotropic particle radiation from the 
central part is not so strong. Since the effect can be misinterpreted
as an evidence for the QGP formation, it should be subtracted from 
the analysis of experimental data.
  
The presence of spectators, which absorb hadrons early emitted in the
direction of normal flow, affects also the development of elliptic 
flow. Particularly, it leads to the creation of the dip in $v_2(y)$ 
in midrapidity range. As expected from simple geometrical 
considerations, the effect is stronger in peripheral collisions.

The transition of elliptic flow from the out-of-plane to in-plane
direction with the rise of energy from 1{\it A} GeV to 11.6{\it A} 
GeV can also be linked to change in geometry of colliding system.
The Lorentz-contracted spectators, which rapidly fly away, provide
more free space for the in-plane development of the flow than 
almost noncontracted nuclei, see Figs.~\ref{fig9} and \ref{fig10}. 

\begin{figure}[htp]
\centerline{\epsfysize=18cm \epsfbox{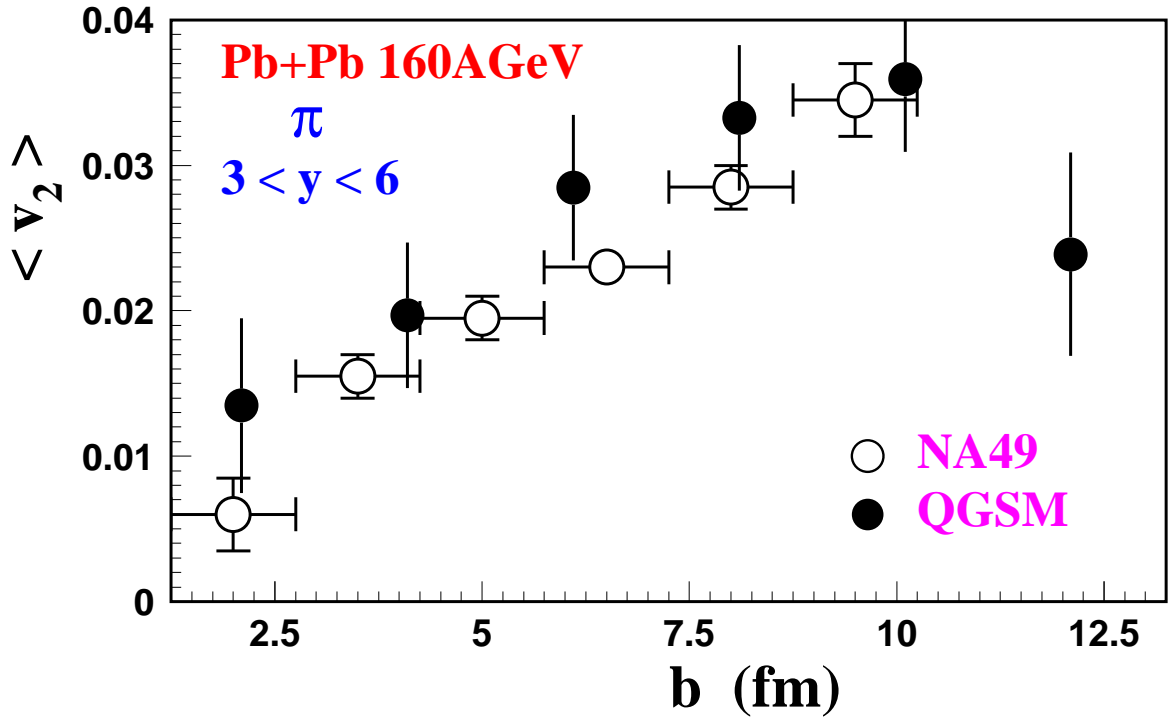}}
\caption{Elliptic flow of charged pions as function of impact 
parameter in rapidity range $3 < y < 6$ in Pb+Pb collisions at 
160{\it A} GeV. Open circles denote the experimental data from
\protect\cite{na49v2}, full circles are the model predictions.
}
\label{fig11}
\end{figure}

It is worth to mention that the elliptic flow of nucleons as a 
function of impact parameter becomes more flat with rising energy
of the collision, while the maximum in $v_2^\pi(b)$ distribution is
shifted to very peripheral events. This tendency is clearly seen in
Fig.~\ref{fig8}. Figure~\ref{fig11} presents the comparison of the
model simulations of the elliptic flow of charged pions in 
$3 < y < 6$ in Pb+Pb collisions at SPS energies with the 
experimental data \cite{na49v2}. We see that the QGSM provides a good
quantitative agreement with the experiment. 
Note that the behaviour of the elliptic flow of charged particles is
determined by the proton elliptic flow at energies below 11.6{\it A}
GeV and by the pionic elliptic flow at 160{\it A} GeV. It means that
although the physics of rescattering in the QGSM is the same in 
peripheral and central collisions, the nuclear matter undergoes a 
transition from a baryon dominated to a meson dominated matter with
rising energy of colliding nuclei. The transition is similar to the
predicted in \cite{PoVo99} transition from hadronic to partonic
degrees of freedom.

\section{Conclusions}
\label{sec6}

The directed and elliptic flow of hadrons in heavy-ion collisions is 
very sensitive to the EOS of the nuclear medium. At low and 
intermediate energies (SIS) hadrons are the relevant degrees of
freedom, and the intranuclear interactions, i.e. the mean field,
determine the EOS as well as the reaction dynamics. With increasing
energy new degrees of freedom are extended, and the formation of 
small domains of a QGP phase might happen already at the SPS energies 
or even below. 
Accompanied by the phase transition to the hadronic phase this 
enforces a softening of the EOS due to the dropping pressure. Thus, 
the disappearance of the directed flow in midrapidity range can be 
considered as an indication of a new state of matter. This conclusion 
is supported by hydrodynamic simulations. No deviations of the 
nucleon directed flow from the straight line in $|\tilde{y}| \leq 1$ 
range have been found in the one-fluid calculations with a pure 
hadronic EOS \cite{Cs94,CsRo99}.  

On the other hand, several microscopic models, which do not explicitly
imply the QGP formation, predict larger or smaller 
deviations of the directed flow from the straight line behaviour
\cite{Br95,BZFF99,LPX99,Sn99} which is presented at low and
intermediate energies. These deviations are attributed
to the shadowing effect, which plays a decisive role in the 
competition between normal flow and antiflow in (semi)peripheral 
ultrarelativistic collisions of nuclei. Hadrons, emitted with small
rapidities at the onset of the collision in the antiflow area can
propagate freely, while their counterparts will be absorbed by the 
flying massive spectators. 

The signal becomes stronger with the rise of the impact parameter. In
collisions with the same impact parameter the antiflow starts to
dominate over the normal flow in the midrapidity range as the reaction
becomes more energetic, i.e. the spectators are more 
Lorentz-contracted and more hadrons can be emitted unscreened with 
small rapidities in the direction of antiflow. Therefore, this effect
should appear in (semi)central collisions with $b \leq 3$ fm at
RHIC energies, and can imitate the softening of the EOS of hot and
dense nuclear matter. However, the disappearance of directed flow
due to shadowing is more distinct for light systems, like S+S or 
Ca+Ca, colliding with the same reduced impact parameter. 
In the case of a plasma creation the effect should be more pronounced 
in large systems like Pb+Pb. Thus, one can distinguish between
the two phenomena, shadowing and quark-hadron phase transition, by
the comparison of the directed flow of nucleons in the midrapidity 
range in light and heavy-ion collisions.
  
The elliptic flow of nucleons and pions is found to change its 
orientation from out-of-plane at 1{\it A} GeV to in-plane at
11.6{\it A} GeV. Since the dynamics of rescattering is the same, the 
effect can be explained by purely geometric reasons, such as stronger
Lorentz-contraction of colliding nuclei. At higher colliding energies
the contracted spectators leave the reaction zone faster, thus giving 
space for the growth of elliptic flow in the reaction plane. 

Results of the simulations appear to favour a similarity of hadron
rescattering in central and peripheral heavy-ion collisions at
energies up to 160{\it A} GeV. QGSM predicts that the $\langle 
v_2^\pi(b) \rangle $-distribution in Pb+Pb collisions at SPS energies 
increases as the reaction becomes more peripheral, in accord with the
experimental data. The elliptic flow of pions in this reaction drops 
only for highly peripheral collisions somewhere at $b \approx 12$ fm.
However, if the data will show the further rise of elliptic flow
even at such impact parameters, this can be taken as an indication
for new processes not included in present version of the model.
The situation awaits better data on both directed and elliptic
flow in the midrapidity range and in very peripheral collisions of
light and heavy nuclei at ultrarelativistic energies.

{\bf Acknowledgements.} 
We are thankful to L. Csernai, E. Shuryak, H. Sorge, H. St\"ocker, 
S.~Voloshin, and Nu Xu for the fruitful discussions and comments.
This work was supported in part by the Bundesministerium f\"ur 
Bildung und Forschung (BMBF) under contract 06T\"U887.

\widetext

\newpage

\newpage
\mediumtext

\begin{table}
\caption{ 
The slope parameters of the directed flow of nucleons 
and pions in light and heavy systems in the midrapidity range at
the SIS (1{\it A} GeV), AGS (11.6{\it A} GeV), and SPS 
(160{\it A} GeV) energies.
}

\begin{tabular}{cccccccc}
  {\rm System} & {\rm Particle} & \multicolumn{6}{c} 
{\rm Impact parameter, $\tilde{b}=b/b_{\rm max}$} \\
\cline{3-8}
  &  & 0.15 & 0.30 & 0.45 & 0.60 & 0.75 & 0.90 \\
\tableline\tableline
S+S (SIS)   & N     & 0.12 & 0.18 & 0.17 & 0.14 & 0.08 & 0.05 \\
            & $\pi$ & 0.00 &-0.02 &-0.03 &-0.03 &-0.02 &-0.01 \\
Au+Au (SIS) & N     & 0.27 & 0.37 & 0.37 & 0.33 & 0.26 & 0.22 \\
            & $\pi$ & 0.00 &-0.03 &-0.05 &-0.08 &-0.10 &-0.07 \\
S+S (AGS)   & N     & 0.06 & 0.10 & 0.13 & 0.13 &-0.02 &-0.05 \\
            & $\pi$ & 0.00 &-0.06 &-0.06 &-0.05 &-0.11 &-0.11 \\
Au+Au (AGS) & N     & 0.18 & 0.30 & 0.34 & 0.31 & 0.11 &-0.05 \\
            & $\pi$ &-0.07 &-0.09 &-0.10 &-0.11 &-0.13 &-0.15 \\
S+S (SPS)   & N     & 0.01 & 0.00 & 0.00 &-0.01 &-0.06 &-0.08 \\
            & $\pi$ & 0.00 &-0.04 &-0.05 &-0.06 &-0.08 &-0.09 \\
Pb+Pb (SPS) & N     & 0.15 & 0.18 & 0.24 & 0.19 &-0.01 &-0.13 \\
            & $\pi$ &-0.03 &-0.04 &-0.08 &-0.09 &-0.10 &-0.16 \\
\end{tabular}
\label{tab1}
\end{table}

\end{document}